# A holographic optical tweezers module for the International Space Station

J. Shane[1][*][a], R. Serati[a], H. Masterson[a], S. Serati[a]

Boulder Nonlinear Systems, Inc., 450 Courtney Way, Lafayette, CO, USA 80026

## ABSTRACT

The International Space Station (ISS) is an unparalleled laboratory for studying colloidal suspensions in microgravity. The first colloidal experiments on the ISS involved passive observation of suspended particles, and current experiments are now capable of observation under controlled environmental conditions; for example, under heating or under externally applied magnetic or electric fields. Here, we describe the design of a holographic optical tweezers (HOT) module for the ISS, with the goal of giving ISS researchers the ability to actively control 3D arrangements of particles, allowing them to initialize and perform repeatable experiments. We discuss the design's modifications to the basic HOT module hardware to allow for operation in a high-vibration, microgravity environment. We also discuss the module's planned particle tracking and routing capabilities, which will enable the module to remotely perform pre-programmed colloidal and biological experiments. The HOT module's capabilities can be expanded or upgraded through software alone, providing a unique platform for optical trapping researchers to test new tweezing beam configurations and routines in microgravity.

## 1. INTRODUCTION

The International Space Station (ISS) National Laboratory has allowed researchers worldwide to expand the frontiers of microgravity research. Among the many fields that benefit from the unique research environment of low earth orbit, researchers of fluids and colloids can study effects such as diffusion-controlled heat and mass transfer, which are normally masked by sedimentation, stratification, convection, and other gravity-related phenomena. Applications of this research include improvement of manufacturing processes and improvement of product shelf life.

The Fluids Integrated Rack (FIR) is part of the US Destiny science module, and is a platform for housing and operating fluids-related research equipment. Since its installation in 2009, the FIR has housed the Light Microscopy Module (LMM), a heavily modified upright Leica DMRAX2 microscope designed as a remotely-controlled multi-user imaging tool. The LMM has permanent modules (such as a doubled YAG for fluorescence microscopy, a lamp for brightfield illumination, multiple cameras, and a planned confocal imaging module), as well as interchangeable sample cells.

The LMM is designed to operate as autonomously as possible, with the completely motorized microscope conducting and recording pre-programmed experiments. The LMM operates while stowed in the wall of the Destiny module; crew intervention is largely limited to changing sample cells, while limited data rates and periodic loss of signal prevents full ground control of the LMM. Ground-based principal investigators (PIs) design experiments and sample cells, and receive camera and video results once experiments are complete.

Our goal was to expand LMM capabilities for a wide range of PIs by designing a holographic optical tweezers (HOT) module that would work with all existing sample cells as well as with other LMM functions such as confocal microscopy. The HOT module is designed to give researchers the ability to actively control 3D arrangements of particles, yet perform this particle manipulation and other measurements without real-time user intervention, through the use of particle tracking. Here, we describe the motivations and desired capabilities for the HOT module, as well as our preliminary design for integration with the LMM.

---

[1] jshane@bnonlinear.com; phone 1-(303)-604-0077; bnonlinear.com

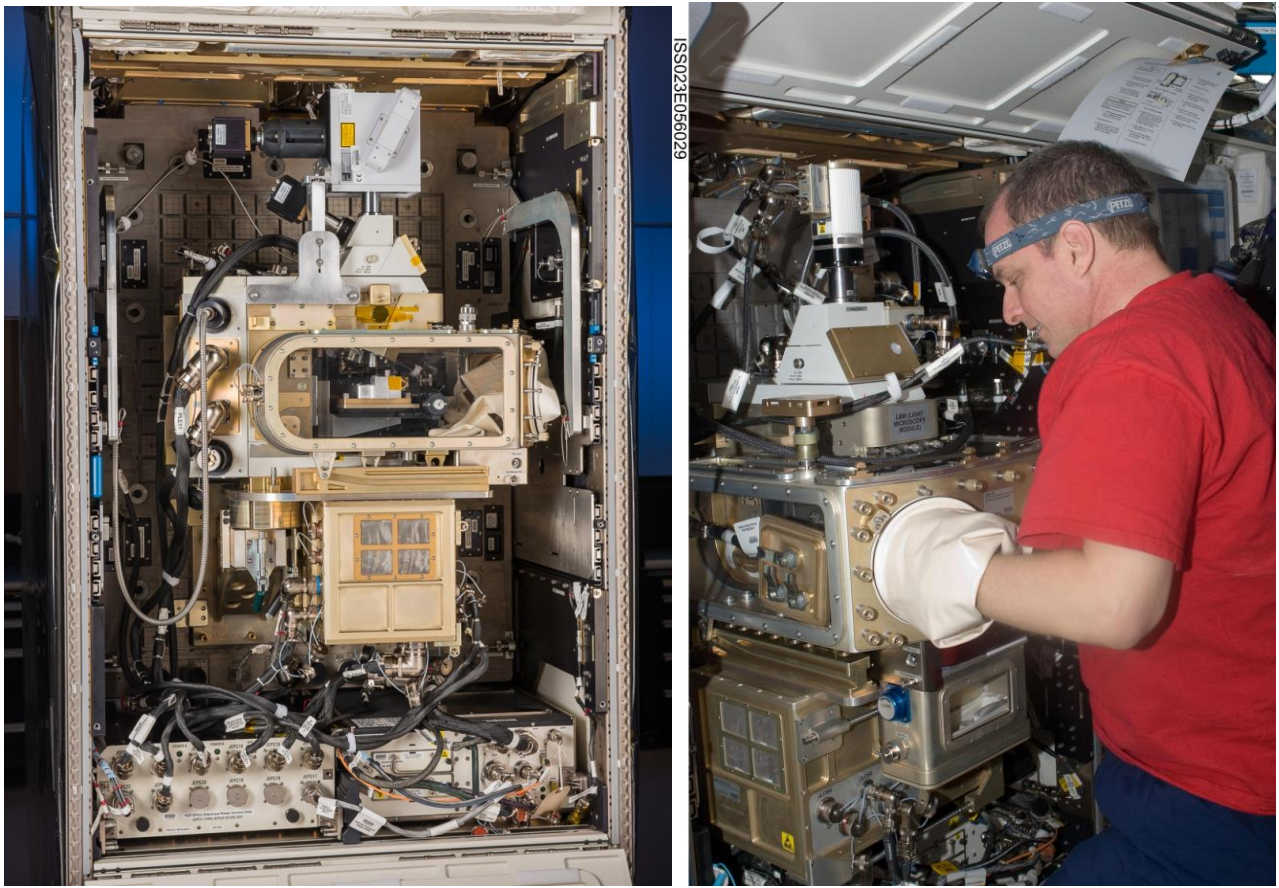

Figure 1. The Light Microscopy Module (LMM), installed in the Fluids Integrated Rack (FIR). Left: The LMM is shown stowed sideways in the FIR (ground-based replica), in the position in which it usually operates autonomously behind a closed door. Right: The LMM in use on-orbit by NASA astronaut T.J. Creamer (confocal microscope not installed). Image credit: Left: ZIN Technologies and NASA Glenn Research Center. Right: NASA.

## 2. MOTIVATION

### 2.1 Active initialization of repeatable colloidal experiments

The first colloidal experiments on the LMM involved passive observation of suspended particles (Advanced Colloids Experiment – Microscopy). Current experiments are now capable of observation under controlled environmental conditions; for example, under heating or under externally applied magnetic or electric fields (Advanced Colloids Experiment-Heated, -Temperature, -Electric field). The addition of a holographic optical tweezers (HOT) [1,2] module to the LMM would give ISS researchers the ability to control precise 3D arrangements of hundreds of particles in parallel, allowing them to initialize and perform repeatable experiments.

### 2.2 Measurement of fluid, cell, and colloidal crystal properties

In addition to formation of custom particle arrangements, research groups worldwide routinely use optical tweezer systems for measuring the properties of fluids, cells, and colloidal crystals, including:

- Linear and nonlinear viscoelastic properties of fluids[3]
- Cell membrane deformability and stiffness (an indicator of cellular health)[4,5]
- Response of colloidal crystals to linear and rotational application of force[2]

With the existence of many mature options for automated particle detection, particle tracking can be combined with the automated configuration of arrays of tweezing beams[6] to allow the above measurements to be completely automated. The HOT module will perform the user's selected or pre-programmed measurement routine without need for step-by-step intervention.

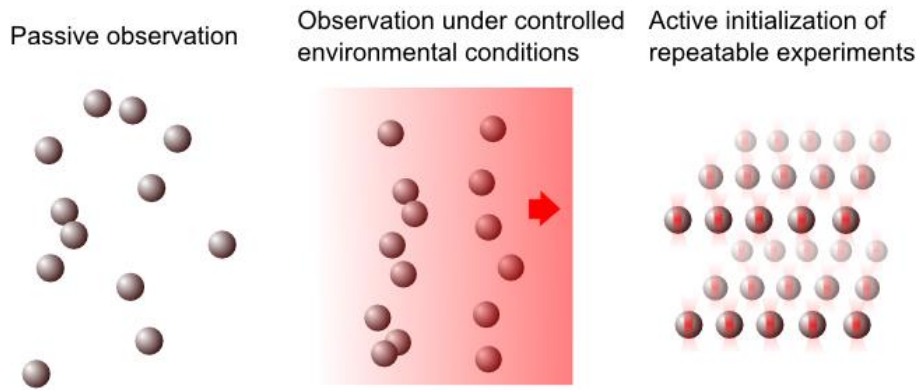

Figure 2. Progression of ISS colloidal research from passive observation to active control.

**2.3 State of the art for 3D micromanipulation: holographic optical tweezing**

In holographic optical micromanipulation, a single input beam reflects off a spatial light modulator (SLM) on its way to the focal plane of a microscope[1]. The SLM has an array of individually-addressable pixels, each of which can be used to modify the phase and/or amplitude of the input beam, producing a highly-controlled 3D light pattern at the focal plane. Through processes such as radiation pressure and scattering, this light pattern can apply force to microscale particles, making the SLM-controlled light pattern a flexible micromanipulation device.

One classic application of a holographically-controlled laser beam is holographic optical tweezing (HOT), also called holographic optical trapping. In HOT, the single input beam is transformed into a set of hundreds of independently-steerable focused beams, each of which is capable of moving microscale particles in 3D using optical gradient and scattering forces. HOT systems have been used to arrange particles in arrays or other arbitrary patterns[7], to control colloidal crystallization dynamics, or even to control individual dislocations in crystal lattices[2]. The shape of each trapping beam can be optimized for different particle types, or to impart angular momentum[1,8,9] to the structure. With the use of graphical processing units (GPUs) for calculations[10–12] and overdrive for speeding up liquid crystal transitions[13], holograms can be calculated and updated in real time.

The use of a liquid crystal on silicon (LCoS) SLM for HOT offers the greatest flexibility as a micromanipulation tool, with the ability to create hundreds of tweezing beams in 3D (as opposed to tens of 2D traps possible with an acousto-optic (AO) deflector to time-share the beam). An LCoS SLM can perform its beam shaping through phase change alone, as opposed to through the amplitude modulation of a digital mirror device (DMD) or AO deflector. This leads to more efficient use of the laser trapping source, as well as the ability to apply phase-based beam shapes and adaptive optics/wavefront correction[14]. Finally, an LCoS SLM can create beams with complex phase patterns, such as helical or Bessel[15,16] beams, allowing them to apply rotational forces and complex wavefront correction in a way that deformable mirror, DMD, and AO deflector devices cannot.

Although HOT is primarily used in a manually-operated laboratory setting, the technique has several features that should make it well-suited for remote-controlled on-orbit use:

- **No moving parts**. Beamsteering and focus change are accomplished entirely by applying voltage to a motionless liquid crystal device.

- **Maintainable and upgradeable via ground-based software alone**. Techniques such as Shack-Hartmann wavefront correction are routinely used to measure and compensate for HOT system aberrations and misalignments, based solely on images returned from the camera. New functionality can be added to the ground-based user interface without changing the on-orbit software.

- **Particle recognition makes automated operation possible.** Centroid-based algorithms can locate particles to within a fraction of their diameter, while trajectory planning and collision avoidance can be used to automatically move these particles to their desired locations[6].

- **Robust and vibration resistant.** Optical tweezers, including holographic optical tweezers, are routinely used as outreach tools, or as trade show demonstrations, operating on non-floating tables in the presence of significant vibration from foot traffic[17].

In the following section, we detail our design for a HOT module for the LMM, which incorporates the optics and electronics necessary to provide full 3D control of hundreds of automated tweezing beams.

## 3. LMM HOT MODULE DESIGN

### 3.1 Tweezing beam introduction strategy

The current and planned LMM configurations rely on epi rather than trans imaging, since their sample cells are usually opaque, so the tweezing beam should be introduced into the epi illumination path. Many of the existing epi ports are used by other modules such as confocal imaging (Figure 3, right) or brightfield epi illumination, while a large fluid-containment system known as the Auxiliary Fluids Container (AFC, Figure 3) restricts access to the area near and below the nosepiece.

Keeping these restrictions in mind, we developed an approach toward integrating HOT with the LMM without compromising LMM capabilities. We focused our design efforts on introducing the tweezing beam and illumination through the LMM's side port (highlighted in green in Figure 3, right), which maximizes the space available for optics and electronics, while avoiding any impact on existing LMM functionality, or on the space available for samples. With this design, imaging can be performed using the existing epi or confocal cameras, and there is no need for trans illumination.

The only physical parts affected are the confocal brace and mounting bracket (highlighted in red and blue in Figure 3, used to counter the torque introduced by the confocal module's spinning disk); these can be remade to provide a mounting location for the HOT module while preserving their mechanical performance by maintaining at least their current size and mass.

The HOT module also requires a specialized filter cube to be mounted in the LMM's filter wheel. Entry through the side port requires the use of two adjacent filter cube slots, one for the specialized side-entry filter cube, and one adjacent empty slot to provide a clear path from the side port to the filter cube. The side-entry filter cube blocks the LMM's epi brightfield illumination, so as described in Section 3.4, we provide brightfield LED illumination in our design.

Figure 4 shows an overview of the LMM with the HOT module attached, showing the proposed location and approximate size of the module. The major components, discussed in more detail in the following sections, are the HOT laser, the SLM, and an illumination LED.

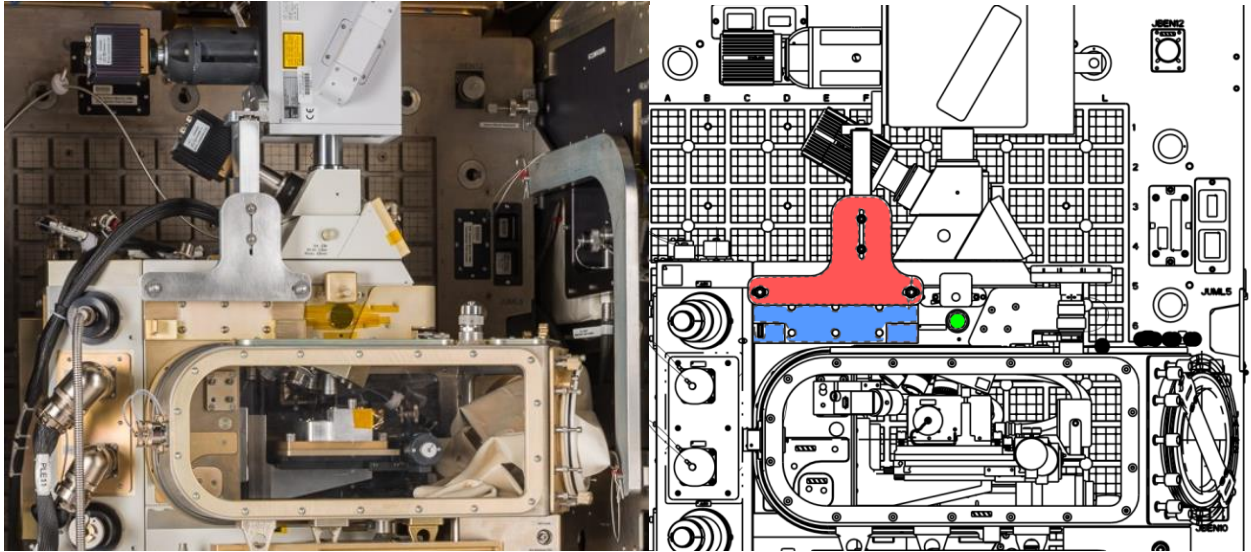

Figure 3. Left: Photograph and right: Drawing of LMM with AFC and confocal module shown. The HOT module optomechanics will be mounted to the brackets shown in red and blue. The confocal brace (shown in red) will be modified to allow the HOT module to be mounted in this location without reducing the brace's mass or minimum extent, maintaining its structural function. The shaped trapping beam will enter through the side port highlighted in green. Image credit: Left: ZIN Technologies and NASA Glenn Research Center. Right: adapted from ZIN Technologies.

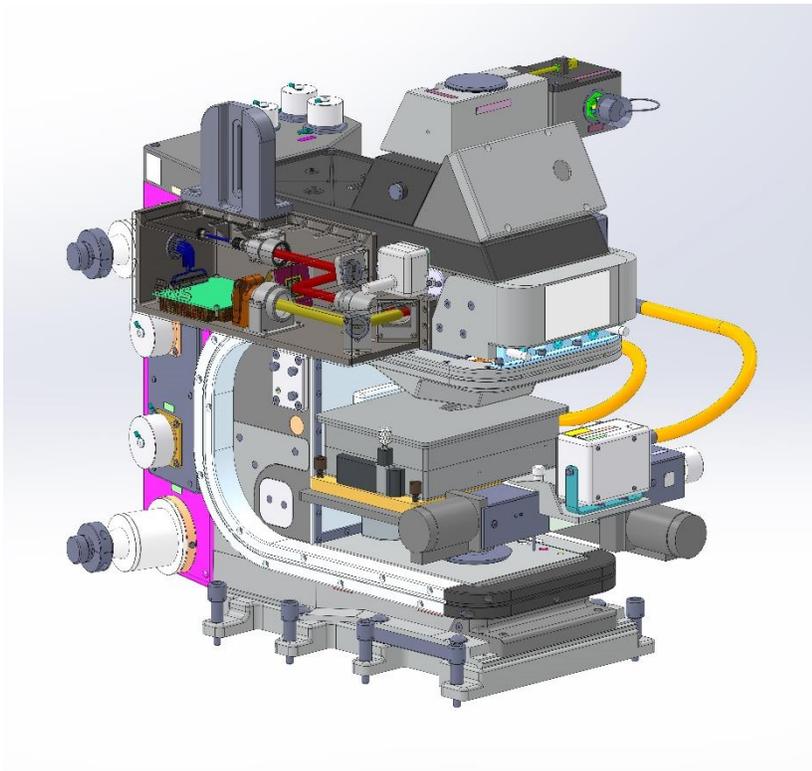

Figure 4. LMM overview with HOT module attached, showing the proposed location and approximate size of the module. The confocal unit is not shown, although the modified confocal brace is included. LMM model: ZIN Technologies. HOT module model: Boulder Nonlinear Systems.

## 3.2 Choice of laser and LED wavelength

During Phase I we chose the design wavelengths for the HOT laser and the LED illumination, keeping in mind availability of parts as well as compatibility with other LMM operations.

The LMM's cameras have silicon-based sensors, which are sensitive to visible wavelengths as well as to infrared wavelengths as long as 1100nm (Figure 5). The LMM's confocal imaging system uses a narrowband 532nm laser (doubled YAG) as illumination, and detects longer-wavelength fluorescence from excited samples (a typical fluorescence emission spectrum from a 532nm-absorbing fluorescent dye is shown in Figure 5). By selecting near infrared (NIR) wavelengths for the HOT module's laser and LED brightfield illumination, we take advantage of an unused wavelength range of the silicon camera, and avoid conflict with the confocal fluorescence imaging.

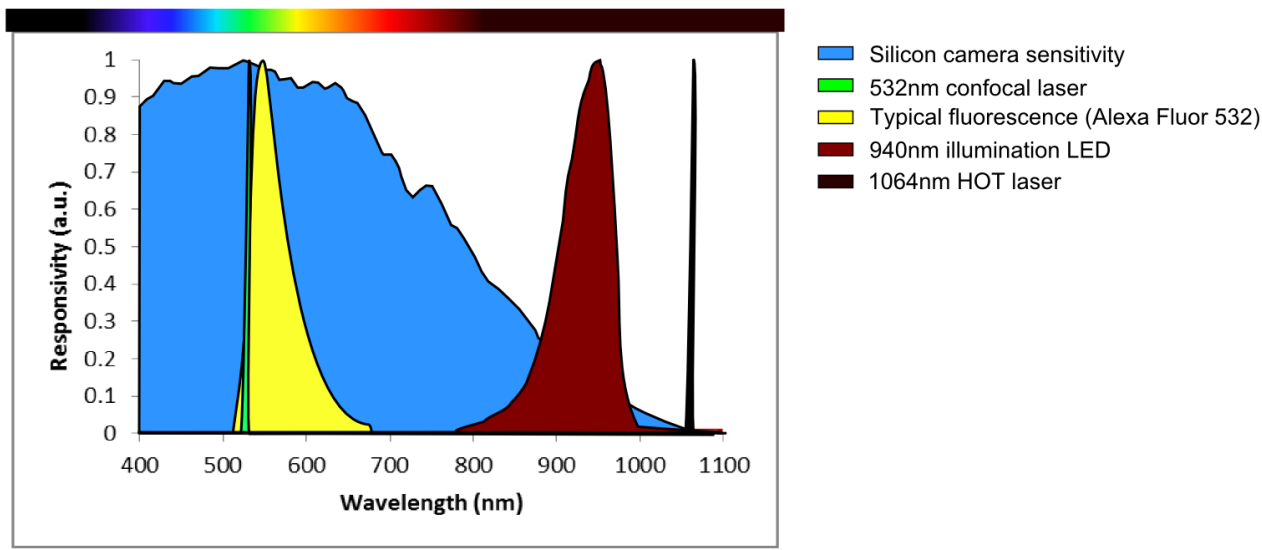

Figure 5. Sensitivity of a typical silicon camera sensor (blue), shown with existing confocal illumination (green) and fluorescence (yellow) ranges, as well as proposed HOT LED illumination (red) and tweezing (black) wavelengths. HOT illumination and tweezing wavelengths are designed to be in the near infrared region, to avoid any conflict with confocal and visible-wavelength imaging. The 940nm LED spectrum shown is that of a Thorlabs M940L3.

To be capable of forming ~300 simultaneous 3D tweezing beams, the tweezing laser needs to be capable of emitting an average power of 3-5W, and this output must be fiber-coupled due to space constraints near the microscope side port. A high-quality single-mode output is also important for forming the tightly focused tweezing beams, and the output should also be linearly polarized for compatibility with the SLM. The choice of 1064nm for the tweezing wavelength maximizes the range of available sources (including fiber lasers, or diode-pumped lasers coupled to single-mode polarization-maintaining fibers). In addition to suitability for the primary application of colloidal micromanipulation, 1064nm excitation also allows for trapping live cells with minimal damage, as long as the cell is trapped for less than about a minute [18,19].

## 3.3 Hardware design overview

Figure 6 shows the HOT module optomechanics in more detail, including the strategies used for introducing the 1064nm HOT laser beam and the 940nm LED illumination. In brief, the HOT beam (red) is introduced via a fiber laser collimator (C1) and steered onto the SLM. The SLM is imaged onto the back aperture of the microscope objective (OBJ) using two lenses, one before the microscope side port (L1), and one attached to the modified side-entry filter cube (L2). At the intermediate image plane, the beam impinges on a mirrored zero order beam block (ZOBB), which deflects any undiffracted light from reaching the sample. The LED illumination (yellow) is collected by a condenser lens (C2) and imaged onto the back focal plane of the objective. A 1064nm notch mirror (DM1) allows 940nm LED illumination to pass through to the side port, while also serving as a steering mirror for the 1064nm HOT beam. The filter cube contains a dichroic filter (DM2) with a cutoff wavelength of 940nm, reflecting nearly all of the HOT beam down through the objective to the sample plane, while acting as a 50/50 beamsplitter for the 940nm illumination.

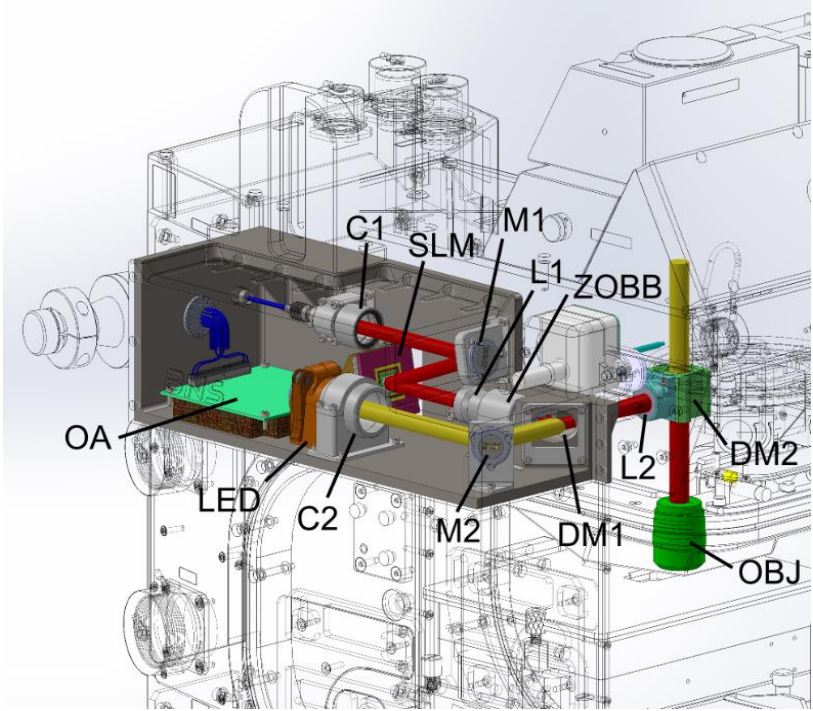

Figure 6. Overview of the HOT module optics. HOT beam (red) is introduced via a fiber laser collimator (C1) and steered onto the SLM. The SLM is imaged onto the back aperture of the microscope objective (OBJ) using two lenses, one before the microscope side port (L1), and one attached to the modified side-entry filter cube (L2). At the intermediate image plane, the beam impinges on a mirrored zero order beam block (ZOBB), which deflects any undiffracted light from reaching the sample. The LED illumination (yellow) is collected by a condenser lens (C2) and imaged onto the back focal plane of the objective. A 1064nm notch mirror (DM1) allows 940nm LED illumination to pass through to the side port, while also serving as a steering mirror for the 1064nm HOT beam. The filter cube contains a dichroic filter (DM2) with a cutoff wavelength of 940nm, reflecting nearly all of the HOT beam down through the objective to the sample plane, while acting as a 50/50 beamsplitter for the 940nm illumination.

### 3.4 LED illumination

One HOT module feature necessitated by side port entry is the addition of a brightfield illumination source. Although the LMM already has an epi light source, illumination from this source will be blocked during HOT operation by the side-entry filter cube. Epi imaging from the confocal module can take place concurrently with HOT module operation, but relies on fluorescence from samples excited at 532nm and emitting at longer wavelengths, which limits confocal imaging to certain fluorescent samples. For maximum user flexibility, the ability to image nonfluorescent samples is important.

To provide brightfield illumination for the HOT module, we couple in light from an LED, whose incoherent light provides widefield illumination without coherence-related speckling or artifacts. This brightfield illumination can be used independently of the HOT module, providing a narrowband IR illumination alternative to the existing broadband visible epi-illumination.

Figure 7 shows the Zemax simulated performance of our illumination design, showing the propagation of LED illumination from the LED to the sample plane. A Zemax intensity plot at the right shows the simulated irradiance at the sample plane, achieving good illumination uniformity over the visual field.

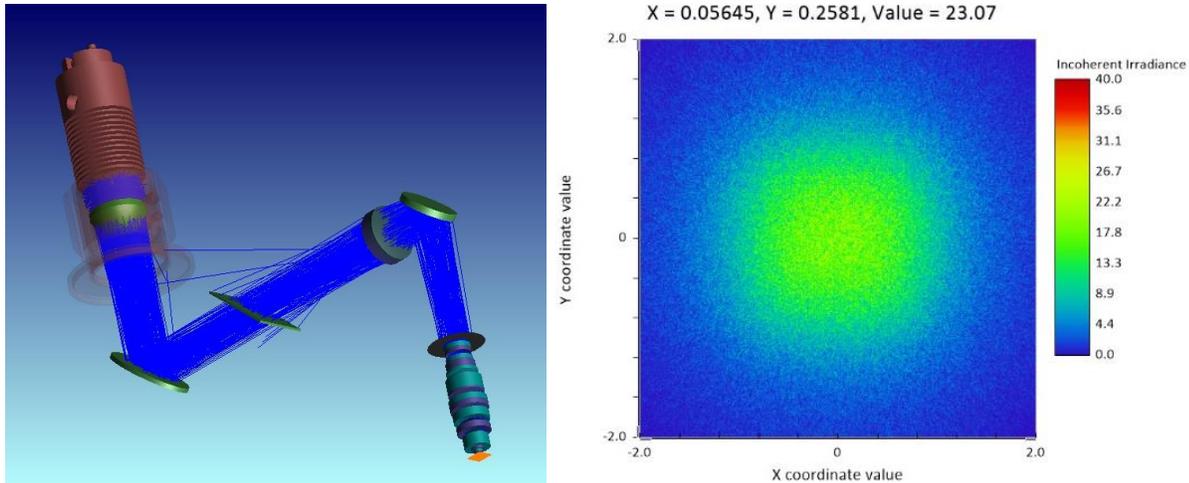

Figure 7. Left: Zemax simulation of the LED illumination propagation from LED to microscope objective focus, showing collection of the LED's emitted radiation by the collimation lens. Right: Zemax simulation of the illumination pattern at the microscope focus, showing good illumination uniformity.

### 3.5 Software

The software libraries that control BNS SLMs, as well as those that perform hologram calculation and aberration correction, are completely modular, allowing them to be incorporated into wrapper programs written in a wide variety of languages, and to operate in both Windows and Linux environments. This modular approach will allow the HOT module's on-orbit SLM control and ground-based user interface to be incorporated with existing LMM/FIR software and communication systems.

Figure 8 shows the concept for communication between the components of the HOT module and the FIR/LMM. Users interact with a ground-based user interface, which allows them to select a predefined preprogrammed automated routine, or to program an automated routine of their own. These commands are then sent to the ISS, where the HOT module's CPU directs the execution of the programmed measurement routines, controlling the SLM, and communicating with the FIR's camera and stage control.

An example of a preprogrammed routine might be the formation of a 3D array of particles, with the user defining the particle locations, and the on-orbit HOT module CPU's automated routine using particle tracking to steer tweezing beams to move particles to the desired locations. To perform such an experiment, the HOT module will (Figure 9):

- Illuminate with LED brightfield
- Monitor the image from one of the LMM's cameras
- Identify the locations of randomly diffusing particles in the image
- Calculate a hologram that will place a tweezing beam at the location of each particle
- Turn on the tweezing laser, with power adjusted as necessary
- Route each particle to its desired location, making sure to avoid collisions between particles
- Check that each particle is now being held in its desired location
- Turn off the tweezing laser, allowing the particles to diffuse once more

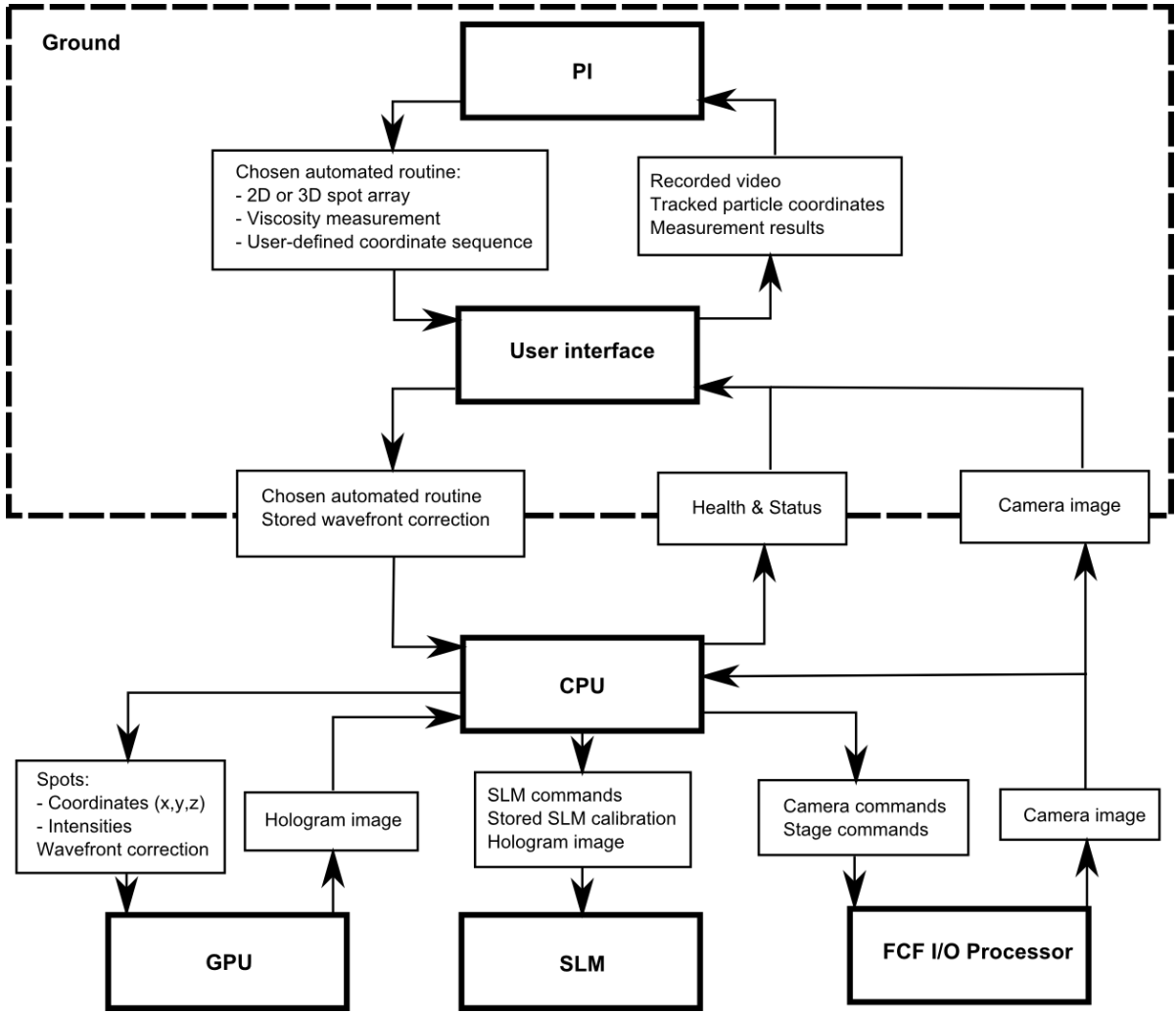

Figure 8. Communication diagram showing the flow of data between different components of the HOT module and FIR

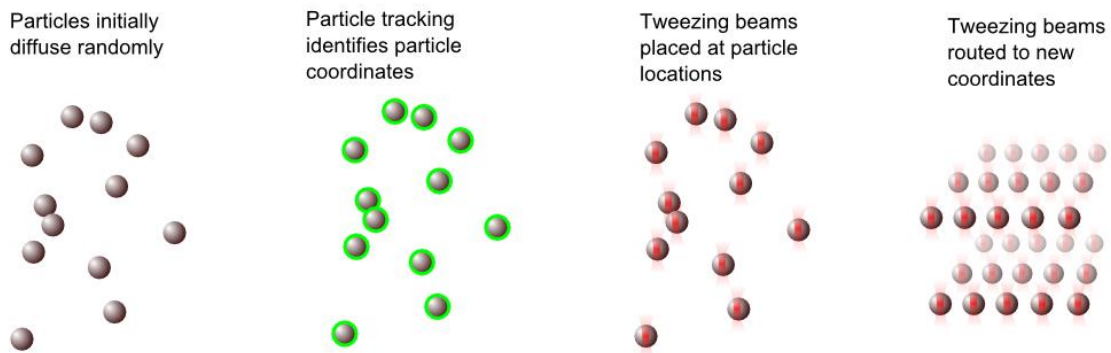

Figure 9. Major automated steps in the formation of a 3D trapped particle array

### 3.6 Spatial Light Modulator (SLM) with driver board, ribbon cable, PCIe cable, and power supply

The SLM system consists of three major components: the driver board, the op-amp board, and the SLM head. All three are shown in Figure 10. The CPU communicates with the driver board via a PCIe host adapter, sending each new hologram image to the driver board via PCIe cable and receiving acknowledgement at the completion of each download. A 32-channel cable transfers data to the op-amp board, which prepares the pixel data for being loaded row by row onto the SLM pixels. A flex cable connects the op-amp board to the SLM head, which houses the liquid crystal pixels that apply phase retardance to the optical beam. The driver board, the largest producer of heat, will be located on the optics bench rear with access to the cooling air flow provided by the optics bench. The length of the flex cable is limited by communications restrictions, so the op-amp board will be located near the LMM side port along with the SLM head and steering optics,

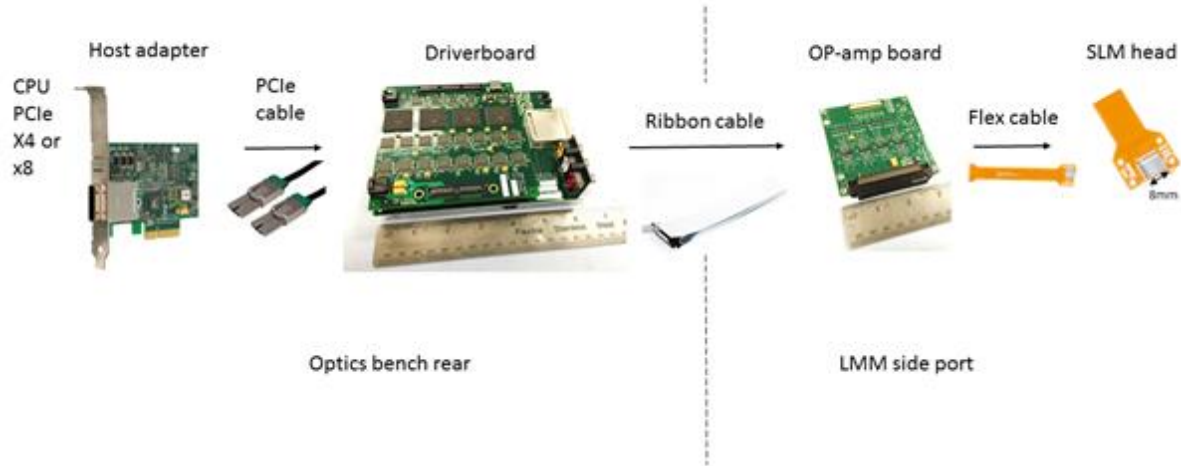

Figure 10. The three major components of the SLM system. The driver board will be located on the optics bench rear, with a 3 meter 32-channel data cable, configured for LMM form factors, passing through to the optics bench front. The Op-amp board and SLM head are located within the HOT module enclosure at the LMM side port.

### 3.7 Electronics and Cabling

The HOT module's optical components are designed to bolt to the LMM side port, while the supporting electronics (CPU with integrated GPU for fast hologram calculation, PCIe host adapter, SLM driver board, and laser head) will be located at the rear of the optics bench as an Avionics Package (Figure 11). A generic package interface (GPI) connector or cable pass-through will supply the needed laser fiber input, SLM data and power, and LED power to the module from the Avionics Package. The length of the cable, at least 3 meters, will allow for the LMM's 90 degree motion between crew use and stowed/operating positions (Figure 1). A custom power supply will convert the ISS 28V DC power to inputs usable by HOT module electronics.

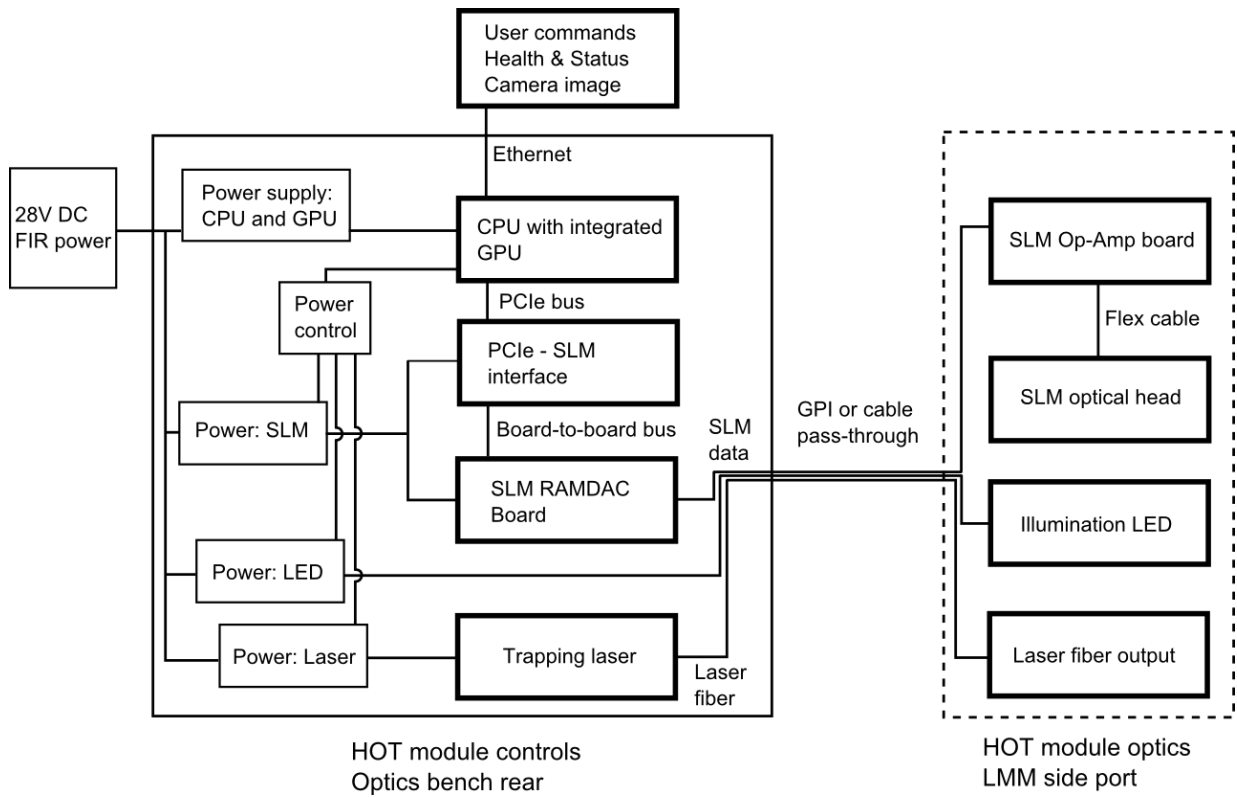

Figure 11. Block diagram showing the components and cabling of the HOT module

## 4. ADAPTATION TO ISS ENVIRONMENT

In addition to hardware and software interfacing considerations for the LMM/FIR, the HOT module will need to be adapted to the environmental conditions of the ISS. Below we discuss some of the most important environmental considerations: microgravity, and the vibration environment.

### 4.1 Microgravity

Although the optical tweezing process itself is not expected to change noticeably in microgravity, the HOT module hardware must be designed for microgravity operation.

In microgravity, heat dissipation cannot rely on the usual mechanism of convective air cooling, since the convective rising of warm air requires gravity – in microgravity, a heat-producing part would remain surrounded by a mostly-stationary region of warm air. Instead, heat dissipation in microgravity requires fans to actively move warm air, or else conductive heat transfer to an actively-cooled surface. In the HOT module design, we avoid the use of fans, which contribute noise and vibration to the ISS environment, as well as torque. We rely instead on conductive heat transfer through the use of heat pads and metal-to-metal connection for those components that generate some degree of heat (the CPU, the SLM op-amp board, the LED, and to a lesser extent the SLM head). Conduction will provide a heat path to a surface entrained in the cooling air flow provided by the FIR facility on the front and the back of the Optics Bench.

Any user-adjustable parts in microgravity should be designed to avoid loose parts, for example, through the use of knobs and captive screws rather than through ordinary screws. The HOT module is designed with no user-adjustable parts – the optical components are motionless and enclosed, with the module activated and controlled through software alone. Because filter cubes and support brackets will have to be replaced during installation, the HOT module is designed for ground-based rather than on-orbit installation, to take place during the LMM's next return to Earth for upgrade.

## 4.2 Vibration

The LMM is a high-vibration environment, due to the effects of machinery and crew movement. The FIR itself has been measured to have vibration amplitude of a few hundred nm at 5Hz, mostly from fans, with the new confocal module not expected to add any more vibration. Tweezers systems operate readily in high-vibration environments, for example as outreach demonstrations, on non-floated tables in areas with high foot traffic. For a rough estimate of the HOT module's ability to maintain trapping in the presence of environmental vibration, we can compare the relative velocity of the fluid during vibration to a typical particle escape velocity. Treating vibration as a sinusoidal movement of the trapping beam with respect to the trapping medium, and considering the worst-case scenario where all vibration is in phase at 5Hz, then the maximum particle velocity during vibration is about 15μm/s. This is about 7x less than the escape velocity of 106μm/s measured by Baek and colleagues [20] for a 1.79μm particle trapped at 1064nm with 16.5mW power. The ISS vibration environment, therefore, is not expected to prevent particle trapping, although it may restrict the size range of trapped particles, and reduce the velocity with which particles can be steered. Optomechanical design, however, will need to be robust enough to survive the extreme vibrations of launch without misalignment.

## 5. CONCLUSION

Here, we have presented a design for a Holographic Optical Tweezers (HOT) module for the Light Microscopy Module (LMM) currently located in the Fluids Integrated Rack (FIR) of the International Space Station (ISS). We have detailed a beam introduction strategy, a concept for mounting the HOT module optics on the side of the LMM, and outlined software, cabling, and support electronics housing. This design could be modified to fit future LMM versions as well.

## 6. ACKNOWLEDGEMENTS

This work was supported by a Phase I SBIR grant from NASA. We would like to thank ZIN Technologies, particularly Dr. Christopher Lant, for many valuable discussions and for information on the current and planned LMM and FIR configurations. We would also like to thank Dr. William Meyer and Dr. Ronald Sicker of NASA Glenn Research Center for their extensive support of this project.